\def\etal{{\it et al.}}

\documentclass[aps,prb,superscriptaddress,a4paper,twocolumn]{revtex4-1}

\usepackage{amsmath}
\usepackage{color}
\usepackage{floatrow}
\usepackage{amssymb}
\usepackage{graphicx}
\usepackage{xcolor}
\usepackage{blindtext}

\begin{document}
\title{Neural-network model for force prediction in multi-principal-element alloys}

\author{R. Singh}
\affiliation{Ames Laboratory, U.S. Department of Energy, Iowa State University, Ames, Iowa 50011 USA}
\author{P. Singh}\email{psingh84@ameslab.gov}
\affiliation{Ames Laboratory, U.S. Department of Energy, Iowa State University, Ames, Iowa 50011 USA}
\author{A. Sharma}
\affiliation{Ames Laboratory, U.S. Department of Energy, Iowa State University, Ames, Iowa 50011 USA}
\affiliation{Sandvik Coromant R\&D, Stockholm, 12679 Sweden}
\author{O.R. Bingol}
\affiliation{Mechanical Engineering, Iowa State University, Ames, Iowa 50011 USA}
\author{A. Balu}
\affiliation{Mechanical Engineering, Iowa State University, Ames, Iowa 50011 USA}
\author{G. Balasubramanian}
\affiliation{Mechanical Engineering \& Mechanics, Lehigh University, Bethlehem, PA 18015 USA}
\author{A. Krishnamurthy}
\affiliation{Mechanical Engineering, Iowa State University, Ames, Iowa 50011 USA}
\author{S. Sarkar}\email{soumiks@iastate.edu}
\affiliation{Mechanical Engineering, Iowa State University, Ames, Iowa 50011 USA}
\author{Duane D. Johnson}\email{ddj@iastate.edu, ddj@ameslab.gov}
\affiliation{Ames Laboratory, U.S. Department of Energy, Iowa State University, Ames, Iowa 50011 USA}
\affiliation{Materials Science \& Engineering, Iowa State University, Ames, Iowa 50011 USA}

\begin{abstract}
\vspace{0.2in}
Atomistic simulations can provide useful insights into the physical properties of multi-principal-element alloys. However, classical potentials mostly fail to capture key quantum (electronic-structure) effects. We present a deep 3D convolutional neural network (3D CNN) based framework combined with a voxelization technique to design interatomic potentials for chemically complex alloys. We highlight the performance of the 3D CNN model and its efficacy in computing potentials using the medium-entropy alloy TaNbMo. In order to provide insights into the effect of voxel resolution, we implemented two approaches based on the inner and outer bounding boxes. An efficient 3D CNN model, which is as accurate as the density-functional theory (DFT) approach, for calculating potentials will provide a promising schema for accurate atomistic simulations of structure and dynamics of general multi-principle element alloys.
\end{abstract}

\maketitle

\section*{Introduction}
Multi-principal-element alloys, often called high- and medium-entropy alloys (HEAs/MEAs), continue to draw significant attention due to their remarkable mechanical behavior.\cite{Cantor2004,Yeh2004,Li2004,Gludovatz20014,George2019,Yang2018,Jo2017,Varvenne2017,Granberg2016,Miracle2017,Ikeda2019,Tsai2014,Singh2018,Singh2020_Vac,Singh2020,Singh2018_2} However, the HEA design space is impractically large to explore mechanical properties of all possible compositions using experiments or first-principles calculations due to limitations on system size. On the other hand, computational modeling by atomistic simulations, especially for the sub-micron length processes, has not been progressed due to the absence of well-defined force fields to describe interatomic interactions.~\cite{AS2016,Sharma012017,Sharma022017} An interatomic potential function refers to the mathematical equation that provides a direct functional relationship between the configurations, positions, and potential energy of a group of atoms. Analytical potentials provide a simpler, direct, closed-form relation between the molecular configurations and their potential energy, which facilitates a quick energy calculation, but they are usually derived by introducing physical approximations.\cite{Artrith2016} These potentials represent a necessary compromise between efficiency and accuracy, given that the essential characteristics of the atomic interactions are reasonably described. In molecular-dynamics simulations,\cite{Plimpton1995} force-field functions, such as the embedded atom method (EAM),\cite{Zhou2004} are available for a limited set of elemental-combinations in HEAs and can be employed to describe the self and cross-interactions between different participating species.

Atomistic simulations that utilize suitable interatomic potentials to describe the atomic interactions can easily overcome the shortcomings of first-principles theory and connect directly to experiments. However, there are not many attempts to address issues related to inter-atomic potentials for multi-principal element alloys such as HEAs.\cite{Grabowski2019} Notably, the absence of a general scheme provides an opportunity for designing reliable, useful, and robust interatomic potential functions. Machine-learning models can provide one such avenue for the accelerated design of interatomic potentials. Although extensively explored using machine learning, the design of interatomic potentials for atomistic simulations was mostly focused on simple elemental or binary systems.\cite{Artrith2016, Liu2017, Behler2016,13,14,16,18,19,20,21,22,Cusentino2020} Also, the volumetric nature of the atomic feature descriptors (i.e., position of atoms, interatomic distances, etc.) makes the machine-learning methods more effective.\cite{26,27,28,31,32,33,35,36,38} If properly trained, a machine-learning-based interatomic potentials can provide the accuracy of density-functional theory (DFT) based methods.\cite{29,33,34} 

Here, we present a deep-learning convolutional neural network (CNN) model seeded with data from density-functional theory (DFT) based {\it ab initio} molecular dynamics (AIMD) simulations to develop interatomic potentials for HEAs. These potentials utilize CNN machine-learning models to construct direct functional correlations between the configuration, atomic position, energy, and forces\cite{Artrith2016} using a consistent set of atomistic data.\cite{Artrith2016,Liu2017,Behler2016} DFT-AIMD-based machine-learned potentials will enable faster evaluation of potentials, especially for large MD or Monte Carlo (MC) simulations general applicability to all possible interactions within a system containing various atomic species, and high-predictive accuracy comparable to DFT methods. These ideas were tested on a ternary TaNbMo MEA within a 54-atom random configurational model, created from Hybrid Cuckoo Search optimized Super-Cell Random Approximates (SCRAPs),\cite{Singh2021}. Although any sized cell could be considered, it is limited by the cost of the DFT data generation. Our 3D CNN model shows great promise for developing alternate approaches for the design of atomistic potentials. 

\begin{figure*}[t!]
\includegraphics[width=0.85\linewidth]{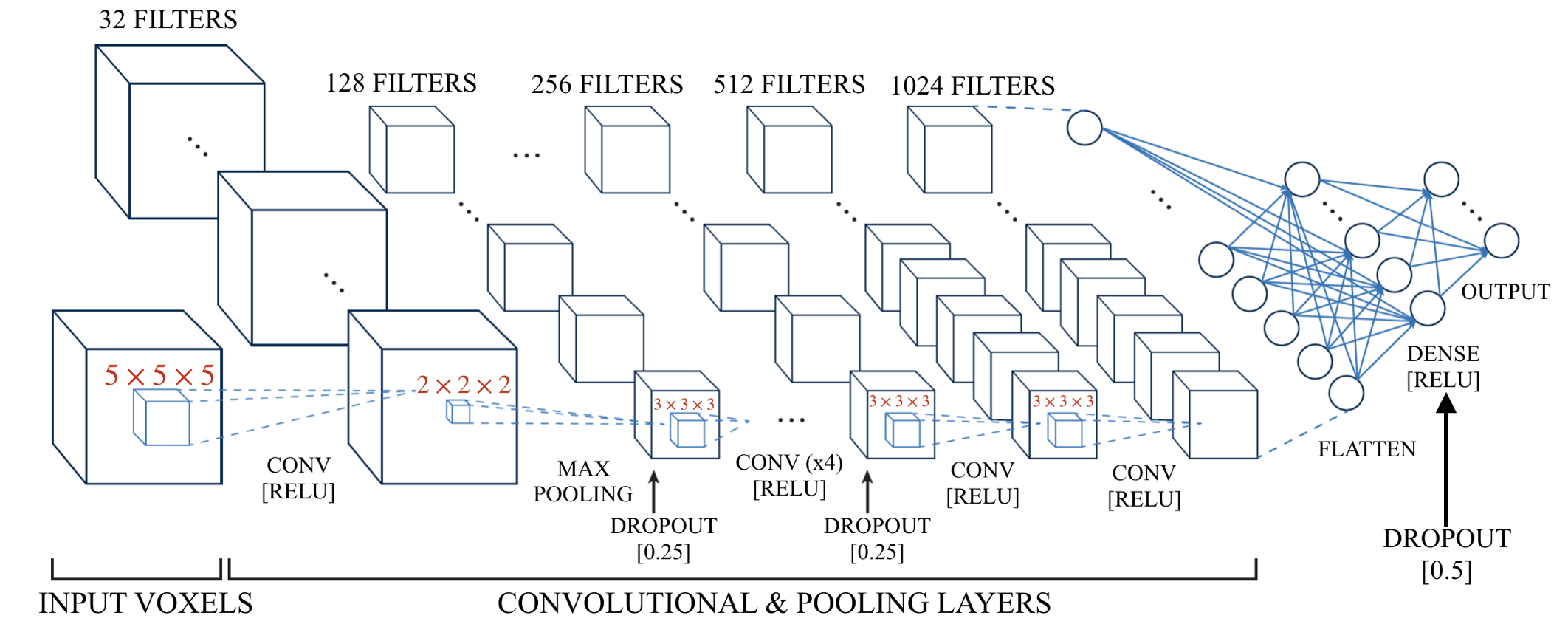}
\caption{Convolutional Neural Network (CNN) architecture was used to learn the relationship between the potential energy surface and the atomic coordinates of the alloy cell. The CNN architecture primarily consists of convolutional layers followed by max-pooling layers and one fully-connected layer at the end succeeded by the output.}
\label{Fig6}
\end{figure*}

\section*{Methods}

\noindent
{\it Deep Convolutional Neural Networks:} Data encountered in real life often involves learning from three-dimensional (3D) data. Naturally, this has been a major area of deep learning research, where 3D CNNs were used to learn from 3D data. One of the earliest works on using 3D CNNs was for object detection (VoxNet) using a 3D voxelized geometry of simple objects (such as a bed or a chair).\cite{wu20153d,ioannidou2017deep} Since then, this idea has been exploited in several areas, such as object detection from point cloud data obtained using LIDARs,\cite{maturana} engineering data used in design for manufacturing applications,\cite{sambitgmp} material microstructures synthesis applications,\cite{singh2018physics,pokuri2018interpretable} and rendering smooth three dimensional graphics.\cite{tatarchenko2017octree} 

A deep neural-network consists of several layers of connections forming one network, which takes a input $x$ and produces an output $y$. Each connecting layer ($l_i$) in the network can be represented as $y_{l_i} = \sigma(W_{l_i} \cdot x_{l_i} + b_{l_i})$, where $\sigma(...)$ represents a non-linear activation function, $W_{l_i}$ and $b_{l_i}$ are the weights and biases, respectively, for connecting the input neurons ($x_{l_i}$) to the output ($y_{l_i}$) neurons. The connections could be as simple as a dense connection between every input neuron and output neuron in the layer. However, all connections in a dense connection layer may not be meaningful and the sample complexity to learn the connections would be high. A convolution connection instead of a dense connection helps in alleviating this issue. The convolution operation ($\otimes$) is given by
\begin{eqnarray} \label{LAMeq}
W[m,n,p] \otimes x[m,n,p] = \nonumber \,\,\,\,\,\,\,\,\,\,\,\,\,\,\,\,\,\,\,\,\,\,\,\,\,\,\,\,\,\,\,\,\,\,\,\,\,\,\,\,\,\,\,\,\,\,\,\,\,\, \\ 
\sum_{i=-h}^{i=h}\sum_{j=-l}^{j=l}\sum_{k=-q}^{k=q}  W[i,j,k] \cdot x[m-i,n-j,p-k]
\end{eqnarray}

A series of convolutional connections, non-linear activations, and pooling forms a CNN. Subsequently, the voxelized data obtained from the process is fed to the CNN model. The network architecture, shown in Fig.~\ref{Fig6}, comprises of multiple convolution layers with max pooling and dropout layers in-between the convolution layers. {Now, we discuss the different non-linear activations used.}

\noindent
{(a)~{\it Activation function 1:}~ The \texttt{LeakyRelu} activation function was used for the convolution layers in the network, which can be defined as:
\begin{equation*}
LeakyRelu(x)=
\begin{cases}
1, x<0\\
\alpha x + 1, x\geq0,
\end{cases}
\end{equation*}
where $\alpha$ is a parameter set by the user. We found that $\alpha = 0.001$ works the best for the current data and network.}\\

\noindent
{(b)~{\it Activation function 2:}~ The \texttt{tanh} function
\begin{equation*}
tanh(x)=\frac{2}{1+e^{-2x}} -1,
\end{equation*}
was used as the activation function in three dense and fully connected layers close to the output of the network.}\\

{Once the CNN has been defined, the weights of each connection is randomly initialized and then optimized using an appropriate loss function.}

\noindent
{(c)~{\it Loss function:}~ The loss function used for training:
\begin{equation*}
l = \frac{1}{|\mathcal{D}|}*\sum_{k\in \mathcal{D}} ({y_{pred}}_k - {y_{true}}_k)^2,
\end{equation*}
where $\mathcal{D}$ is the dataset to learn from and $|\mathcal{D}|$ is the number of data points used for training.}

{With the use of different optimization schemes, we train the CNN and then perform testing accordingly.}

\noindent
{\it DFT-AIMD Calculations}: We utilized ab-initio molecular dynamics (AIMD) simulations as implemented in Vienna Ab-{\it initio} Simulation Package \cite{Kresse1993,Kresse1994} for energy, displacements and force calculations for the ternary TaNbMo MEA in a 54-atom supercell. The Perdew-Burke-Ernzerhof (PBE)\cite{Perdew1996} generalized gradient exchange-correlation potential and the projected-augmented wave (PAW) potentials were used.\cite{Blochl1994,Kresse1999} The 54-atom supercell was generated using SCRAPs method\cite{Singh2021} and was relaxed in the {tt VASP} pseudo-potential code with a 400 eV plane-wave energy cutoff and with a Born-Oppenheimer convergence of 10$^{-6}$ eV at each time step. A Monkhorst-Pack k-mesh of 10$\times$10$\times$10 is used in all AIMD simulations. A Nos{\'e} thermostat is applied at different temperatures (100 to 1500 K) for the various simulated cases, as required by the exchange-correlation functional to initiate atomistic displacements in the TaNbMo.

\begin{figure}[t]
\includegraphics[width=0.8\linewidth]{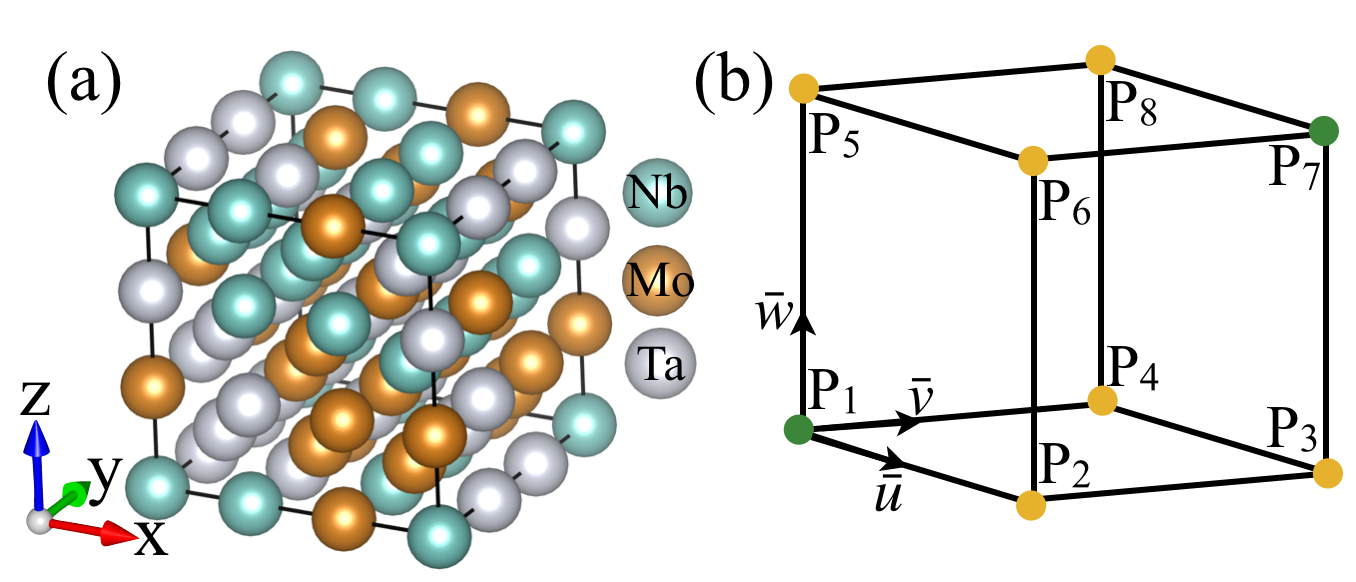}
\caption{(a) Forces from TaNbMo 54-atom model of the random alloy, fully relaxed using DFT, were used to develop potentials. (b) Schematic representation of volume element (voxel) of bcc unit cell based data for voxlization.}  
\label{Fig5}
\end{figure}

The DFT-AIMD simulations were performed on TaNbMo 54-atom ternary MEA model in Fig.~\ref{Fig5}a to extract displacements (\textit{x}, \textit{y}, \textit{z}) and forces (\textit{F$_{x,y,z}$}). A temperature range from 100-1500 K in the steps of every 200 K was considered to include the temperature effect on displacements and forces. The resulting dataset consists of \textit{70k} unique snapshots with relevant displacements, forces, and energies of all atoms in TaNbMo 54-atom SCRAP. A structural descriptor \textit{V}, as expressed in Eq.~\ref{morse_eqn}, was employed to capture the interatomic interactions in TaNbMo. {The \textit{70k} unique snapshots obtained is randomly split into \textit{50k} samples for training, \textit{10k} samples for validation and \textit{10k} samples for testing. This validation strategy enables us to ensure we do not overfit to the training set.}

\begin{figure*}[t]
\includegraphics[scale=0.37]{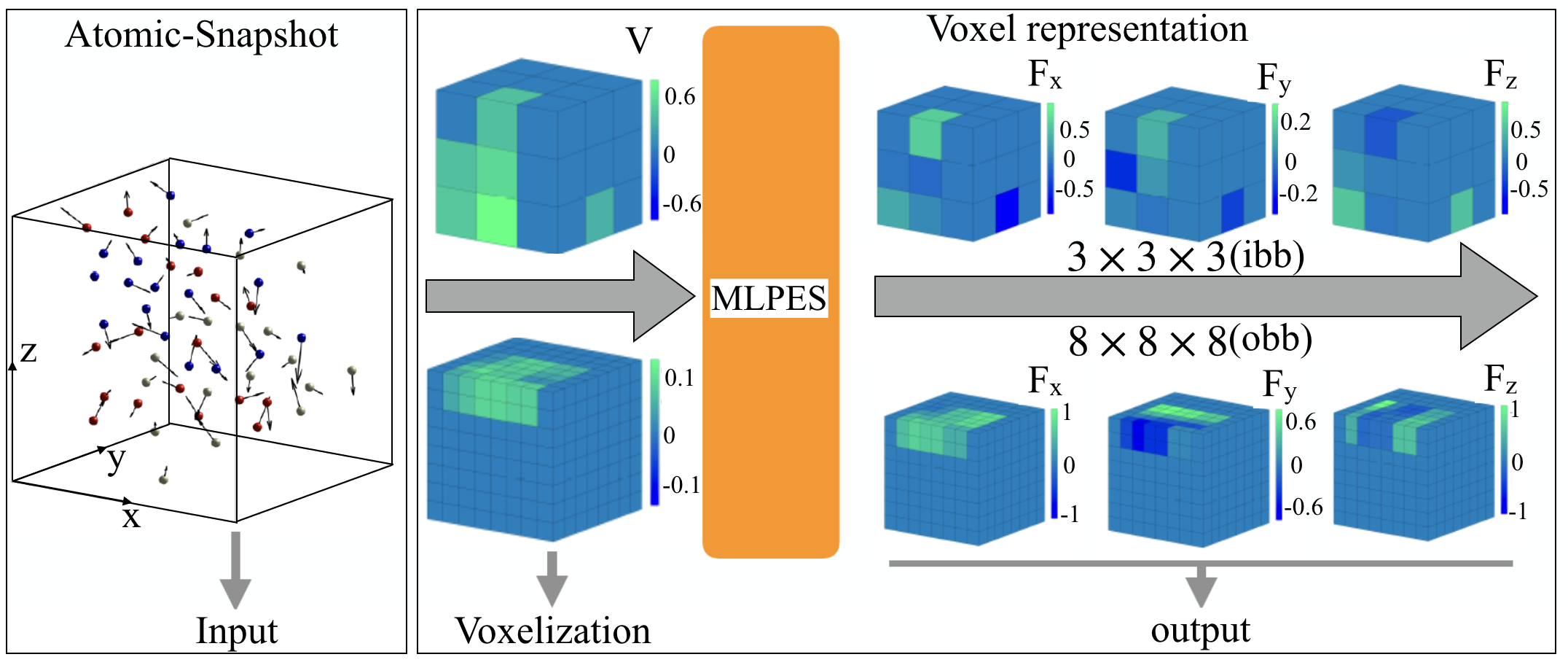}
\caption{(Right-panel) Atomic snapshot and (left-panel) two voxel representations (inner bounding box (ibb) and outer bounding box (obb) ) of 54-atom TaNbMo random alloy are shown. A 3$\times$3$\times$3 (ibb; top left-panel) and 8$\times$8$\times$8 (obb; bottom left-panel) voxel grid were generated to fit forces (\textit{F$_{x,y,z}$}) using neural-network machine-learning model. Both forces (F$_{x,y,z}$) and structure descriptors (V) can be visualized through these voxels.}
\label{Fig1}
\end{figure*}

\noindent
{{\it Voxelization}}: Typical DFT-AIMD output is extensive and complex for direct analysis with a deep-convolutional neural network (DNN) because the DFT-AIMD data structure consists of a large set of atomic coordinates and properties in a multi-component system. Therefore, we used the idea of voxels to represent atomic configuration. The data in \textit{voxel} can be represented in a simple cellular structure bounded by a rectangular box, which is analogous to the \textit{volume element (V)}. The \textit{voxelization} allows to generate a \textit{voxel} grid with a user-specified resolution. Each \textit{voxel} in the grid can contain `zero', one or more atoms depending on their spatial (\textit{x},\textit{y},\textit{z}) coordinates. The \textit{voxel} generation also depends on an inside-outside test that compares the atomic coordinates of the atoms with respect to the grid boundaries of the \textit{voxel}. The grid-bounds are defined by the minimum and maximum points that can form the diagonal of the volume elements. 


We used a \textit{voxel} based representation of the atomic configuration ternary 54 atom TaNbMo MEAs, with:
\begin{subequations}
\begin{align} \label{Eqn:InsideOutside}
0 &< \vec{k} \cdot \vec{u} \leq \vec{u} \cdot \vec{u}  \\
0 &< \vec{k} \cdot \vec{v} \leq \vec{v} \cdot \vec{v}  \\ 
0 &< \vec{k} \cdot \vec{w} \leq \vec{w} \cdot \vec{w} 
\end{align}
\end{subequations}
where $\vec{u} = P_2 - P_1$, $\vec{v} = P_4 - P_1$, and $\vec{w} = P_5 - P_1$ are the basis vectors of the voxel and $\vec{k} = P_v - P_1$. Once the atoms in simulation domain are decomposed to their respective voxels, the atomic properties are computed for each voxel. Here, each voxel contains the structure descriptor \textit{V}, Eq.~\ref{morse_eqn}. In the generalized framework, the user is free to define desired descriptor to account for structural, chemical, electronic or other characteristics. The test in Eq.~\ref{Eqn:InsideOutside} was performed to check whether a point of interest, e.g., \textit{P$_v$}, lies inside or outside the corresponding \textit{voxel}. \textit{P$_v$} is considered inside the voxel if the inequalities in Eq.~\ref{Eqn:InsideOutside} are satisfied.

The voxel grid generation for 54-atom TaNbMo was implemented using two distinct methods i.e., inner bounding box (\textit{ibb}) and outer bounding box (\textit{obb}), as shown in Fig.~\ref{Fig1}. For the \textit{ibb}, the bounding box was computed for each atomic snapshot and created the voxel representation based on that bounding box, which allows each snapshot to have its unique (internal) bounding boxes. Therefore, the invariance in \textit{ibb}) with respect to the scale in atomic locations allows to capture the scale and relative interaction of atoms within this representation. The voxel representation generated with resolution options of (4$\times$4$\times$4), (8$\times$8$\times$8), and (16$\times$16$\times$16) in Fig.~\ref{Fig1}. The resolution of the voxel grid defines the number of volume elements bounded by the grid and, consequently, the size of the processing data. Alternatively, the \textit{obb} approach was used to generate voxel representation of entire set of \textit{70k} snapshots. The \textit{obb} representation is not invariant of the scale nor does it preserves the uniqueness of the atomic locations, therefore, it becomes important to have a very fine resolution of voxel grids. The \textit{obb} also provides a better generalization over \textit{ibb} as it automatically includes the scale variance on the expense of additional computational time. Additionally, the differences in atom locations are attributed to the effect of kinetic energies in the AIMD simulations at the different temperatures by using a proper voxel resolution all possible interactions within the alloy are included in the training dataset. Figure~\ref{Fig1} illustrates a snapshot of coordinates in the 54-atom TaNbMo MEA cell as well as corresponding \textit{ibb} and \textit{obb} representations before and after passing through the CNN framework. 

A multi-processing approach with \texttt{Numba} \cite{Lam2015} was employed to accelerate the voxelization operations. A \texttt{first-in-first-out} (FIFO) \texttt{queue} is created (with queue package) and all \textit{70k} snapshots of displacements and force configurations were added to the \texttt{queue}. The FIFO \texttt{queue} of the configurations was processed using parallel programming within the threading package. The voxelization subroutines are optimized and pre-compiled using the just-in-time compilation capability of \texttt{Numba} package. Applying optimization and pre-compilation steps before the subroutine execution allowed us to substantially reduce (2 to 3$\times$) the processing time spent for generating the voxel grids on a workstation equipped with {\tt Intel Xeon E5-2630 v3} processor and 32 gigabytes of RAM.

{The structural descriptor $V_{(x,y,z)}$ was employed to capture the inter-atomic interactions in HEAs:}
\begin{eqnarray}\label{morse_eqn}
V_{(x,y,z)} =\sum_{i=1}^{N} Z _{i} e^{-[(x-x_{i})^2+(y-y_{i})^2+(z-z_{i})^2]/2\gamma^{2}}.
\end{eqnarray}
{Here, V with voxel coordinates $(x,y,z)$ can be intensity or the value assigned. The (\textit{x},\textit{y},\textit{z}) are the normalized Cartesian coordinates with respect to maximum magnitude, \textit{$Z_i$} refers to the atomic number. Parameter $\gamma$ is the variance, which is typically 2.0 for a normal Gaussian distribution.}

\section*{Results and Discussion}

{The effect of voxel resolution and different voxelization methods on the training of the CNN model is an important aspect of testing the proposed framework. To account for the differences in the resolution of the inner-bounding box (\textit{ibb}) and outer-bounding box (\textit{obb}), we modify the parameters in Eq.~\ref{LAMeq} by reducing the filter size and number of filters during the training of \textit{ibb} and \textit{obb}. For the \textit{obb} method, the deep neural-network consists of layers with 256 filters and the kernel size of 2 $\times$ 2 $\times$ 2 in the first convolution layer. The strides of the convolution filters are 1$\times$1$\times$1 in each convolution layer with a dropout of 0.5 in between the alternate layers. Subsequent layers had a 50\% reduction in filters, while kernel size and strides were constant. In \textit{ibb} model, increasing the filters and their size along with strides had negligible effect on the performance of the model. All the convolution layers have the LeakyReLu activation ($\alpha$ = 0.001) -- except the final 3 dense layers, which had a {\tt tanh} activation. Finally, the entire network is reduced to three fully-connected layers, the layers bearing correspondence to \textit{F$_{x,y,z}$} obtained from the DFT-AIMD. For the present implementation, we use the Adam \cite{kingma2014adam} optimizer for minimizing the loss function. The test data was used for validating the performance of the training, while the training and validation was done for 100 epochs. The analysis were carried out in a computing architecture with two {\tt Nvidia Tesla V100} GPUs.}

\begin{figure}[t]
\includegraphics[width=0.78\textwidth]{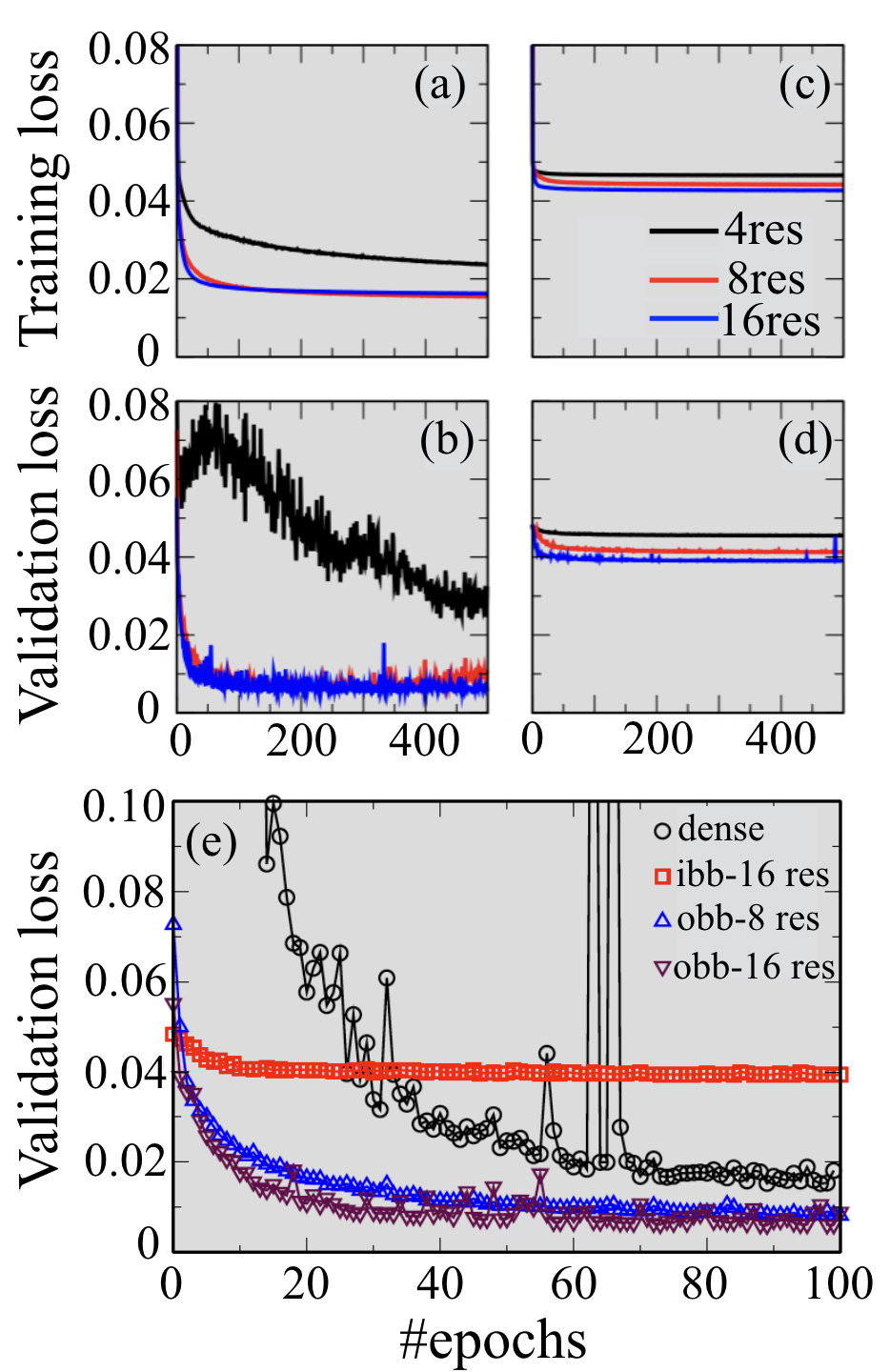}
\hfill
\caption{ (a,c) Training and (b,d) validation loss calculated as function of epochs for outer bounding box (\textit{obb}) and the inner bounding box (\textit{ibb}). The losses are shown on different voxel resolutions of 4$\times$4$\times$4, 8$\times$8$\times$8 and 16$\times$16$\times$16 for \textit{ibb} and \textit{obb} in 54-atom TaNbMo cell. (e) A comparison of validation loss on a dense network against proposed CNN model. Different resolution of voxels were used as an input for \textit{ibb} (16$\times$16$\times$16) and \textit{obb} (8$\times$8$\times$8; 16$\times$16$\times$16).}
\label{Fig2}
\end{figure}


In Fig.~\ref{Fig2}, we have discussed training and validation losses at three resolutions of the voxel grids within \textit{ibb}. For \textit{obb} methods. The results in Fig.~\ref{Fig2}a-d indicates that the \textit{obb} is more effective than the \textit{ibb}. For \textit{obb} in Fig.~\ref{Fig2}a-b, training and validation losses are around 0.02 and 0.01, respectively, at the (8$\times$8$\times$8) voxel resolution. Further increase in resolution only marginally improves the performance (validation) for the \textit{obb}, i.e., the voxel grid for the \textit{obb} needs to be greater than or equal to 8 in order to minimize the losses. Therefore, we set the voxel resolution to 8$\times$8$\times$8 for \textit{obb} for optimum performance.

\begin{figure}[t]
\includegraphics[width=1\linewidth]{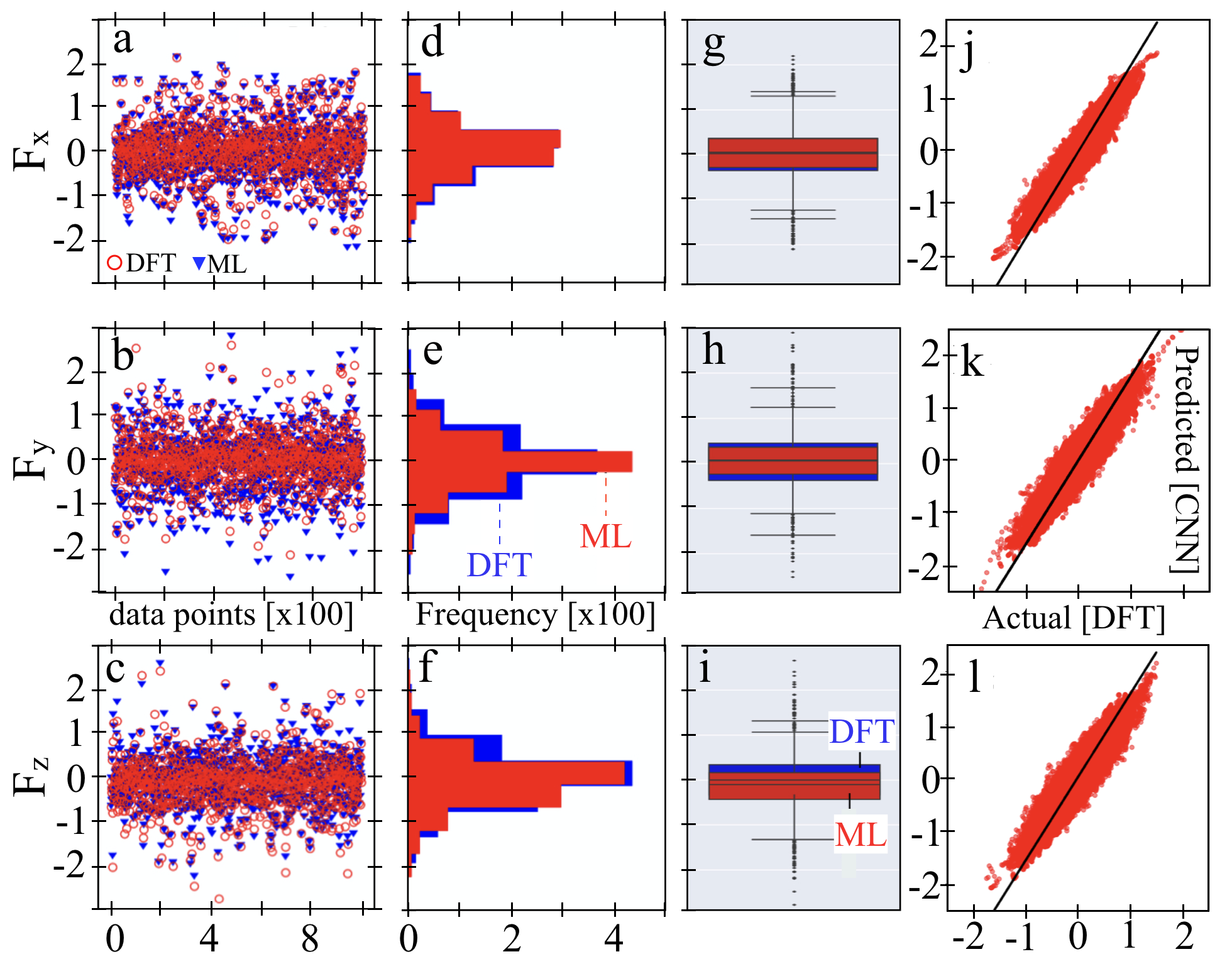}
\caption{The forces from 54 atom cell were fitted using our CNN model on 8 $\times$ 8 $\times$ 8 voxel resolution. (a-c) Spread of the force data (F$_{x,y,z}$) from CNN model with respect to DFT data, (d-f) distribution of F$_{x,y,z}$, and (g-i) the median alongside 25\% and 75\% quantiles spread.{(j-l) The correlation for the 3 forces (F$_{x,y,z}$) with Pearson correlation coefficient of 0.942 is shown using 10,000 data points.}}
\label{Fig4}
\end{figure}

To show the effectiveness of the voxelization framework with respect to a conventional neural network, we plot a comparison of validation losses in Fig.~\ref{Fig2}e at different resolutions. We found that the CNN with \textit{obb} method outperformed the fully connected deep neural network representing a dense configuration. However, for the \textit{ibb} voxelization, the performance was poor in comparison to a full dense configuration, while for the \textit{obb} methodology shows a significant reduction in the losses, implying an enhanced accuracy. {The validation losses for the dense network is around 0.02 while the \textit{obb} at (16$\times$16$\times$16) voxel resolution yields a loss around 0.0075 (loss was defined as a dimensionless quantity scaled by max(F$_{x,y,z}$)).} The inherent nature of the CNN to learn local feature more effectively and perform well even with a higher number of learning parameters can be utilized to develop network architectures to generate the potential energy surface for complex multi-component alloy systems more effectively than the use of traditional dense networks.

The accuracy of the CNN model can be further discussed through a comparison of the DFT-AIMD calculated and CNN-predicted forces in Fig.~\ref{Fig4}a-i. Figure~\ref{Fig4}a-f shows the distribution of the 1000 test data points for DFT-AIMD calculated forces (F$_{x,y,z}$) along along (\textit{x},\textit{y},\textit{z}). The DFT-AIMD calculated F$_{x,y,z}$ (in blue) is largely distributed around the average, i.e., 0, in Fig.~\ref{Fig4}a-c. We can see the CNN predicted F$_{x, y, z}$ in Fig.~\ref{Fig4}a-c are in very close agreement with DFT-AIMD forces as shown by the histogram in Fig.~\ref{Fig4}d-f and box plot in Fig.~\ref{Fig4}g-i. Both DFT-AIMD and CNN datasets show a Gaussian distribution in F$_{x, y, z}$ except for some outliers and correlate very well. The median and the (25, 75)\% quantiles match exactly for F$_{x}$ in Fig.~\ref{Fig4}g while F$_{y}$ and F$_{z}$ show weak deviations for some data points from the DFT-AIMD forces in Fig.~\ref{Fig4}g-i. In the case of F$_{y}$ Fig.~\ref{Fig4}h, the median of DFT-AIMD and CNN force data are significantly close; however, the width of the box does not completely overlap for the two cases. It shows that the model does not best capture the variance in the DFT-AIMD force data for F$_{y}$. A similar deviation was found between the DFT-AIMD and CNN F$_{z}$ data sets in Fig.~\ref{Fig4}i. 

{Fig.~\ref{Fig4}j-l shows a strong positive correlation between the actual (DFT) and predicted (CNN) force values with a Pearson correlation coefficient of 0.942. This further establishes the robustness of the proposed CNN framework. Our framework will provide alternative solutions to the computationally expansive DFT-based simulations by modeling interactions between atoms for designing inter-atomic potential for large-scale calculations. The open-source model and input data are provided online using {\tt GitHub} (will be provided upon acceptance). More functionality for generating MD-ready potentials will be added in the subsequent stages.}

\section*{Conclusion} 

In conclusion, we proposed an efficient framework by combining deep-convolution neural networks with voxelization techniques to learn the potential-energy surface of multi-component alloys, including multi-principal-element (random) alloys. Two voxelization strategies: (a) internal bounding box (\textit{ibb}), and (b) outer bounding box (\textit{obb}) were used to test the efficacy of the proposed framework. We tested our approach on ternary TaNbMo high-entropy alloy. We show that \textit{ibb} captures the critical features of scale and relative interaction of atoms as it considers each configuration as an independent feature. On the contrary, the OBB method defines the bounding box for the entire data set. From the comparison, it is highlighted that the \textit{ibb} method cannot be generalized effectively, while the \textit{obb} method is not invariant of the scale but can offer significant generalization capabilities. Moreover, we also compare our results with a full-fledged dense network and found that the proposed CNN architecture is suited to learn the local features effectively even with a higher number of learning parameters.

The key novelty here is using a volumetric representation of the physical descriptors using a voxel grid and using it to predict the atomistic potential for multi-principal element alloys. To the best of our knowledge, this is the first approach that uses voxel-based (3D) convolutional neural networks (CNNs) for predicting interatomic potential in multi-principal element alloys. Overall, the implementation provides pathways for generating machine-learning (ML) driven frameworks to address the design of potential-energy surfaces for complex multicomponent alloy systems.  

\section*{Acknowledgements} 
Work at Ames Laboratory was funded by the U.S. Department of Energy (DOE), Office of Science, Basic Energy Sciences, Materials Science and Engineering Division. Ames Laboratory is operated for the U.S. DOE by Iowa State University under Contract No. DE-AC02-07CH11358. In part, the work was also supported by the Office of Naval Research (ONR) through awards N00014-16-1-2548, N00014-18-1-2484 and by the U.S. AFOSR under the YIP grant FA9550-17-1-0220 and DARPA-PA-18-02-02 (AIRA) project. Any opinions, findings and conclusions or recommendations expressed in this publication are those of the authors and do not necessarily reflect the views of the sponsoring agencies.

\end{document}